\documentclass[prd,12,twocolumn,groupedaddress,showpacs,nofootinbib]{revtex4}
\pdfoutput=1
\usepackage{graphicx}
\usepackage{bm}
\usepackage{times}
\usepackage{slashed}
\usepackage{color}
\usepackage{color}

\begin{document}

\begin{flushright}
LPT-Orsay-17-07
\end{flushright}

\title{GUT Models at Current and Future Hadron Colliders and Implications to Dark Matter Searches}
\author{Giorgio Arcadi$^1$, Manfred Lindner$^1$, Yann Mambrini$^2$, Mathias Pierre$^2$, Farinaldo S. Queiroz$^1$}

\affiliation{$^{1}$Max-Planck-Institut f\"ur Kernphysik, Saupfercheckweg 1, 69117 Heidelberg, Germany \\
$^2$Laboratoire de Physique Th\'eorique, CNRS, Univ. Paris-Sud, Universit\'e  Paris-Saclay, 91405 Orsay, France}
\begin{abstract}
\noindent 
Grand Unified Theories (GUT) offer an elegant and unified description of electromagnetic, weak and strong interactions at high energy scales. A phenomenological and exciting possibility to grasp GUT is to search for TeV scale observables arising from Abelian groups embedded in GUT constructions. That said, we use dilepton data (ee and $\mu\mu$) that has been proven to be a golden channel for a wide variety of new phenomena expected in theories beyond the Standard Model to probe  GUT-inspired models. Since heavy dilepton resonances feature high signal selection efficiencies and relatively well-understood backgrounds, stringent and reliable bounds can be placed on the mass of the $Z^{\prime}$ gauge boson arising in such theories. In this work, we obtain 95\% C.L. limits on the $Z^{\prime}$
mass for several GUT-models using current and future proton-proton colliders with  $\sqrt{s}= 13~{\rm TeV},\, 33~{\rm TeV},\,{\rm and}\, 100$~TeV, and put them into perspective with dark matter searches in light of the next generation of direct detection experiments.
\\
\begin{center}
{\it Dedicated to the memory of Pierre Bin\'etruy (1955-2017)}
\end{center}

\end{abstract}

\maketitle

\noindent
\section{Introduction} 

Many popular extensions of the Standard Model (SM) rely on the existence of new spin-1 states, possibly with sizable couplings with the SM fermions. This new fields can be promptly interpreted as the gauge bosons of a new $U(1)$ symmetry, spontaneously broken above the Electroweak (EW) scale. An appealing case would be a broken scale not above few TeV, since this would make the new particle, typically dubbed $Z'$~, accessible by collider searches\footnote{This work is focussed on Abelian extensions of the SM. Non-Abelian extensions of the SM are similarly interesting, featuring the presence of both $Z'$ and $W'$ bosons; however they will not be explicitly discussed here.}. 

The simplest way to couple the $Z'$ to SM fermions consists in assuming that the latter are charged under the new symmetry group. Even if this is not the case a coupling with the SM can be provided by a kinetic mixing term~\cite{Babu:1997st} between the Hypercharge field strengh and the one of the new boson. It might be already argued that a natural embedding of this setup is represented by Grand Unified Theories (GUT). Interestingly the minimal viable GUT groups, like e.g. $SO(10)$, have higher rank as the SM group $SU(3)_c \times SU(2)_L \times U(1)_Y$. Under suitable conditions the GUT group can be spontaneously broken, at a first stage, into the SM and an additional U(1) component with the latter, spontaneously broken at some scale above the EW one. As already pointed, a phenomenologically interesting scenario is represented by the case in which the intermediate group is broken to the SM at scales not exceeding few TeVs. 

In case the $Z^{\prime}$ boson can be produced at current colliders, and if the $Z'$ boson possessed  sizable couplings to the SM fermions, it would provide a clear signal represented by resonances in the dilepton, dijet final states peaked at the $Z^{\prime}$ mass.  These searches have been intensively conducted at the LHC and many of the results have been made available \cite{Aad:2012hf,ATLAS:2012ipa,CMS:2013qca,Aad:2014cka}.

Indeed, at the LHC, the high selection efficiencies and relatively small and well understood backgrounds make the dilepton channel a great laboratory probe of new physics at the TeV scale. For these reasons ATLAS and CMS collaborations have reported the most stringent bounds on some $Z^{\prime}$ models that have sizable couplings to charged leptons \cite{Aaboud:2016hmk,Aaboud:2016cth}. Recently ATLAS collaboration has collected at $\sqrt{s}=13$~TeV, $13.3~\mathrm{fb^{-1}}$ of integrated luminosity to exclude $Z^{\prime}$ masses below $3-4$~TeV at 95\% C.L. \cite{ATLAS:2016cyf}. For the Sequential Standard Model, which stands for a $Z^{\prime}$ boson that couples to SM fermions precisely like the Z boson, the lower mass limit of $4.05$~TeV was derived. Additional models were investigated such as the GUT-inspired model $U(1)_{\eta}$, $Z^{\prime}_{\eta}$ for short, which yielded a lower mass limit of $3.43$~TeV.

Motivated by the theoretical importance of these models and the upcoming data collection at the LHC, and future generation of proton-proton colliders, in this letter we cast 95\% C.L. bounds on mass of the $Z^{\prime}$ gauge boson arising in many GUT-inspired models, complementary to previous collider studies. In particular, we will extend the present LHC limits to GUT-inspired models, and also provide projected limits for $\sqrt{s}=13,14,33,100$~TeV, for a variety of integrated luminosities, reaching up to $5\, \mathrm{ab^{-1}}$ in the case of a $100$~TeV collider~\footnote{Our analysis is focused on hadron colliders. Linear $e^+ e^-$ would be also an interesting option~\cite{Richard:2014vfa}.}.  

Moreover, we put our finding into perspective with dark matter endeavors in light of ongoing and next generation of direct detection experiments, namely XENONnT, LZ, and Darwin \cite{Aprile:2012zx,Baudis:2012bc,Aprile:2015uzo,Rizzo:2016yzf,McKinsey:2016xhn,Szydagis:2016few,Aalbers:2016jon} \footnote{There are other important direct detection experiments planned for the future but not particularly sensitive to ours models \cite{Agnes:2014bvk,Agnes:2015ftt,Angloher:2015ewa,Hehn:2016nll,Amole:2016pye,Amole:2017dex}.}. Since both collider and direct dark matter detection observables are dictated by the $Z^{\prime}$ interactions an interesting degree of complementarity between these searches is expected \cite{Frandsen:2012rk,Arcadi:2013qia,Alves:2013tqa,Buchmueller:2014yoa,deSimone:2014pda,Alves:2015pea,Chala:2015ama,Jacques:2016dqz,Fairbairn:2016iuf,Arcadi:2017kky} as we discuss further on.

The paper is organized as follows. In the next section we will describe in more detail the benchmark scenarios adopted in our study. In section III we will discuss our analysis procedure and present the limits
we obtained.  In section IV, we exploit the complementarity between direct dark matter detection and collider searches for $Z^{\prime}$ bosons before
concluding. 


\section{GUT Models}

\begin{table}
\begin{tabular}{|c|c|c|c|c|c|c|}
\hline
  & $Z^\prime_\chi$ & $Z^\prime_\psi$ & $Z^\prime_\eta$ & $Z^\prime_{LR}$ & $Z^\prime_{B-L}$ & $Z^\prime_{SSM}$ \\
\hline
D & $2\sqrt{10}$ & $2\sqrt{6}$ & $2\sqrt{15}$ & $\sqrt{5/3}$ & 1 & 1 \\
\hline
$\hat{\epsilon}^u_L$ & -1 & 1 & -2 & -0.109 & 1/6 & $\frac{1}{2}-\frac{2}{3}\sin^2 \theta_W$ \\
\hline 
$\hat{\epsilon}^d_L$ & -1 & 1 & -2 & -0.109 & 1/6 & $-\frac{1}{2}+\frac{1}{3}\sin^2 \theta_W$ \\
\hline
$\hat{\epsilon}^u_R$ & 1 & -1 & 2 & 0.656 & 1/6 & $-\frac{2}{3}\sin^2 \theta_W$ \\
\hline
$\hat{\epsilon}^d_R$ & -3 & -1 & -1 & -0.874 & 1/6 & $\frac{1}{3}\sin^2 \theta_W$ \\
\hline
$\hat{\epsilon}^\nu_L$ & 3 & 1 & 1 & 0.327 & -1/2 & $\frac{1}{2}$ \\
\hline
$\hat{\epsilon}^l_L$ & 3 & 1 & 1 & 0.327 & -1/2 & $-\frac{1}{2}+\sin^2 \theta_W$ \\
\hline
$\hat{\epsilon}^e_R$ & 1 & -1 & 2 & -0.438 & -1/2 & $\sin^2 \theta_W$ \\
\hline
\end{tabular}
\caption{\footnotesize{Table of couplings of the SSM and GUT-inspired models under investigation.}}
\label{tab:Zpcouplings}
\end{table} 

A Grand Unified Theory (GUT) is a model where the three gauge interactions of the SM which govern the electromagnetic, weak, and strong interactions degenerate  into one value, i.e. an unified interaction. This unified description of these forces is characterized by a larger gauge group, such as $SO(10)$ and $E_6$, spontaneously broken at a scale $M_{\rm GUT}$, typically above $10^{16}\,\mbox{GeV}$
to respect the proton lifetime constraints.

New particles predicted by GUT models are expected to have masses around the GUT scale, thus beyond the reach of any foreseen collider experiments. Nevertheless, signs of of grand unification at high energy scales take place via (for instance) fast proton decay or electric dipole moments of elementary particles \cite{Langacker:1980js}. It is however possible that the breaking of large groups like $SO(10)$ or $E_6$ to the SM gauge groups occurs through different phases, opening the possibility of the existence of states at an intermediate lower scales, possibly accessible to collider experiments. TeV scale manifestations of Grand Unification can be searched via the signal of a $Z^{\prime}$ gauge boson that possesses coupling strength with SM fermions as predicted by GUT constructions. In our work we will consider generic $Z^{'}$ models which correspond to Grand Unification through
$SO(10)$ and $E_6$ symmetry groups as proposed in ~\cite{Langacker:2008yv,Han:2013mra}.
 
$SO(10)$ is a rank-6 group, thus allowing for an extra $U(1)$ component with respect to the SM gauge group. A very natural one is represented by $B-L$, with $B$ and $L$ being respectively the baryon and lepton numbers, as new (spontaneously broken) symmetry. We will consider, in alternative, the case in which the $Z'$ originates from the Left-Right symmetry, which can be described by the following breaking pattern for $SO(10)$~\cite{Langacker:1980js,Hewett:1988xc,Abada:2008gs,Lindner:1996tf,Dueck:2013gca,Awasthi:2013ff,Arbelaez:2013nga,Awasthi:2013ff,Patra:2015bga,Deppisch:2015cua,Deppisch:2016scs,Babu:2016bmy,Hati:2017aez}:$SO(10) \rightarrow SU(3)_C \times SU(2)_L \times U(1)_R \times U(1)_{B-L}$ \cite{Awasthi:2013ff}\footnote{one could also consider $SO(10) \rightarrow SU(3)_C \times SU(2)_L \times SU(2)_R \times U(1)_{B-L}$}. The $U(1)_R \times U(1)_{B-L}$ is then broken to $U(1)_Y$ at a scale $M_{Z'} > M_Z$. The $Z'$ particle relevant for DM phenomenology is a mixture of the gauge bosons of the two $U(1)$ components (as a consequence its coupling with SM fermions rely on a linear combination of their $R$ and $B-L$ charges).

A larger variety of $Z'$ models is based on the $E_6$ gauge group. Indeed, given its higher rank, two extra $U(1)$'s, with respect to the SM gauge group, can be embedded in it. Among the many possible decompositions of $E_6$, two anomaly free gauge groups arise by the following breaking pattern~\cite{Langacker:2008yv}: $E_6 \rightarrow SO(10) \times U(1)_\psi$, $SO(10) \rightarrow SU(5) \times U(1)_\chi$. The $Z'$ associated to the collider phenomenology is, in general, a linear combination of the two components associated to the two $U(1)$'s and schematically expressed as:$Z'=\cos\theta_{E_6}Z'_\chi+\sin\theta_{E_6}Z'_{\psi}$. In this work we will consider three specific assignations for the angle $\theta_{E_6}$: pure $Z'_\chi$ and $Z'_{\psi}$, thus corresponding, respectively, to $\theta_{E_6}=\pi/2$ and $\theta_{E_6}=0$, and a string theory inspired scenario, $Z_\eta=\sqrt{\frac{3}{8}}Z_\chi-\sqrt{\frac{5}{8}} Z_\psi$. Interestingly, the $Z_\chi^\prime$ model features very similar interactions to models based on the $SU(3)_L$ gauge group \cite{Foot:1994ym,Dias:2004dc,Dias:2009au,Profumo:2013sca,Kelso:2014qka,Cogollo:2014jia,Dong:2014wsa,Alves:2016fqe,Lindner:2016bgg}.

As comparison we will also include in our analysis the so called Sequential Standard Model (SSM) consisting in the same assignation as the SM $Z$-boson of the couplings of the $Z'$ with SM fermions.
 
For our phenomenological study we can encode all the considered scenarios in a Lagrangian of the form: 
\begin{equation}
\mathcal{L}=\sum_f g_f \bar f \gamma^\mu \left(\epsilon^f_L P_L + \epsilon^f_R P_R\right) f Z^{'}_\mu
\end{equation} 
where $g_f=g \approx 0.65$ in the case of the SSM and $g_f=g_{\rm GUT}=\sqrt{\frac{5}{3}}g \tan\theta_W \approx 0.46$ for the GUT inspired constructions. 
$\epsilon_L^f (\epsilon_R^f)$ are the couplings associated to the left (right)-handed and their values, for the considered models, are reported in table~\ref{tab:Zpcouplings} (notice that we have used the parametrization $\epsilon_{L,R}^f=\frac{\hat{\epsilon}^f_{L,R}}{D}$ ~\cite{Han:2013mra}.

Notice that we are considering the case where the couplings of the $Z'$ with the SM fermions are determined only by the quantum numbers of the latter with respect to the new symmetry groups. It is nevertheless well known that kinetic mixing between the field strengths of different $U(1)$ gauge bosons is not forbidden neither by Lorentz not by gauge invariant. Furthermore, even once set to zero at the three level, it could be radiatively generated in presence of fermions charged under both the $U(1)$ components. A kinetic mixing term between the $Z'$ and the $Z$ would induce, after EW symmetry breaking mass mixing between the two states and change the couplings of the $Z'$ with respect to the values reported in table~\ref{tab:Zpcouplings}. For simplicity we will assume that kinetic mixing between the $Z'$ and the $Z$ boson is negligible.

Have thus far presented the benchmark models under study we will now move to the collider analysis.

\section{Dilepton Limits} 

Searches for isolated lepton pairs in the final state are considered to be a clean environment to probe new physics at the TeV scale, we will briefly review the reason for such. The most relevant background contributions arise from the Drell-Yan processes. In the dielectron data, Top Quarks, Diboson, Multi-jet and W+jets also subdominantly contribute to the background.  The misidentification of jets as electrons also known as jet-fake rate give rise to the multi-jet and $W$+jets channels as background events. To suppress background from misidentified jets as well as from hadron decays inside jets, electrons are required to satisfy some isolation criteria. It is required that the transverse energy ($E_T$) deposited by the electron to be contained in a cone of size $\Delta R=0.2$. Moreover, events with electrons with pseudo-rapidity ($1.37  <|\eta| < 1.52$) are removed from the analysis because this transition region between the central and forward regions of the calorimeters feature degraded energy resolution. Lastly,  with no charge identification requirement, the electrons have to be isolated, within a cone of size $\Delta R =10\textrm{ GeV}/p_T$, where $p_T$ is the transverse momentum of the electron track. These criteria lead to a selection efficiency of about 87\%, and product of acceptance-efficiency of nearly 70\% for TeV scale dielectron resonances \cite{ATLAS:2016cyf}.

In the dimuon channel, the background from multi-jet and W+jets are irrelevant, since muon misidentification rate is relatively small. In this case only Drell-Yann, Top Quarks, Diboson are important. As for muons opposite-charge assignments are applied. The efficiency is about 94\%, but the product acceptance-efficiency is degraded compared to the dielectron, being about 44\% for TeV scale dimuon resonances \cite{ATLAS:2016cyf}.

Systematic uncertainties in the dielectron channel are at the level of 7\% (25\%) for the signal (background), whereas for the dimuon channel, systematic uncertainties read about 17\% (25\%) for the signal (background). The much larger systematic uncertainties in the dimuon channel are due to uncertainties in the reconstruction efficiency which are in the ballpark of 16\% \cite{ATLAS:2016cyf}. 

That said, to clearly see the power that heavy dilepton resonance searches have at probing new physics we need to also compute the number of signal events. To evaluate the impact of the 13 TeV LHC search for dilepton resonances with $13.3~ \mathrm{fb^{-1}}$ of integrated luminosity \cite{ATLAS:2016cyf}, we simulate the process $pp  \rightarrow Z^{\prime} \rightarrow e^+e^-$ allowing for the presence of jets but requiring the charged leptons to be isolated using MadGraph5 \cite{Alwall:2011uj}, clustering and hadronizing
jets were held within Pythia \cite{Sjostrand:2006za}, and simulating detector effects accounted for with Delphes3 \cite{deFavereau:2013fsa}. We have adopted the CTEQ6L parton distribution functions throughout.

\begin{figure}[!t]
\includegraphics[width=\columnwidth]{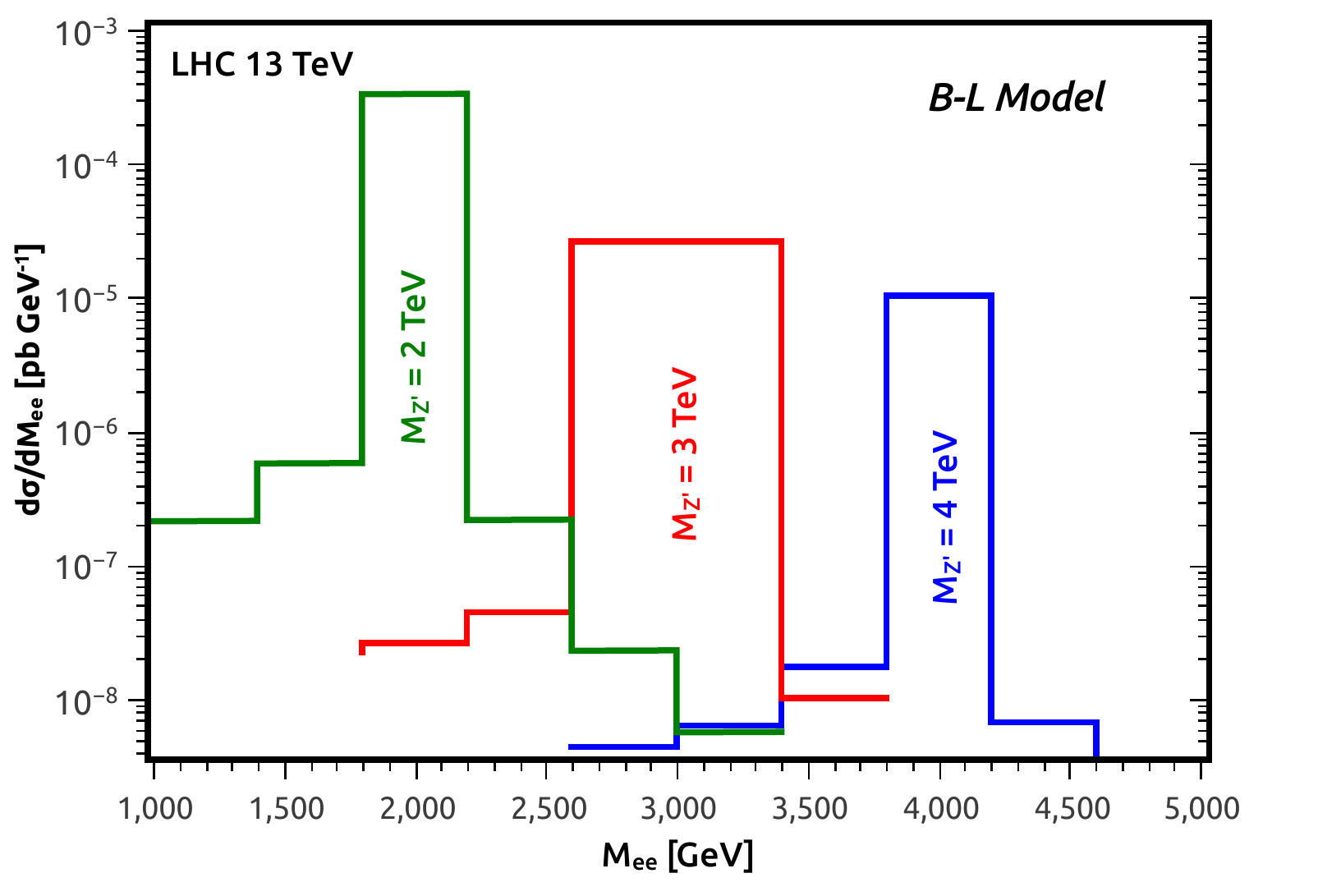}
\caption{Differential cross section for the dielectron channel at 13 TeV. The $p_T$ and $E_T$ cuts given in Sec.\ III were applied to the figure.}
\label{fig:1}
\end{figure}

The signal events were selected with the same criteria used by~\cite{ATLAS:2016cyf} which in summary read:

\begin{itemize}
\item  $E_T(e_1) > 30 \,{\rm GeV}, E_T(e_2) > {\rm 30 \,GeV}, |\eta_e| < 2.5$,
\item  $p_T(\mu_1) > 30 \,{\rm GeV}, p_T(\mu_2) > 30 \,{\rm GeV}, |\eta_{\mu}| < 2.5$,
\item $80 \,{\rm GeV} < M_{ll} < 6000 \,{\rm GeV}$,
\end{itemize}where $M_{ll}$ is the 
invariant mass of the lepton pair used to enhance  signal-to-noise ratio. In the analysis, the presence of a narrow resonance in the dilepton invariant mass has been assumed. In Fig.\ref{fig:1} we show the differential cross section as function of the dilepton invariant mass for the B-L model at $13$~TeV. A narrow resonance is  rather visible. From Fig.\ \ref{fig:1}, one can clearly see the pronounced peak in the dilepton invariant mass coinciding with the mass of the $Z^{\prime}$ gauge boson. A similarly behavior appears also in the remaining models except for the SSM, which has a mildly large decay width. Anyways, we will see further that our results fully agree with ones presented by ATLAS collaboration regarding the SSM. 

\begin{figure*}[t]
\begin{center}
\includegraphics[scale=0.3]{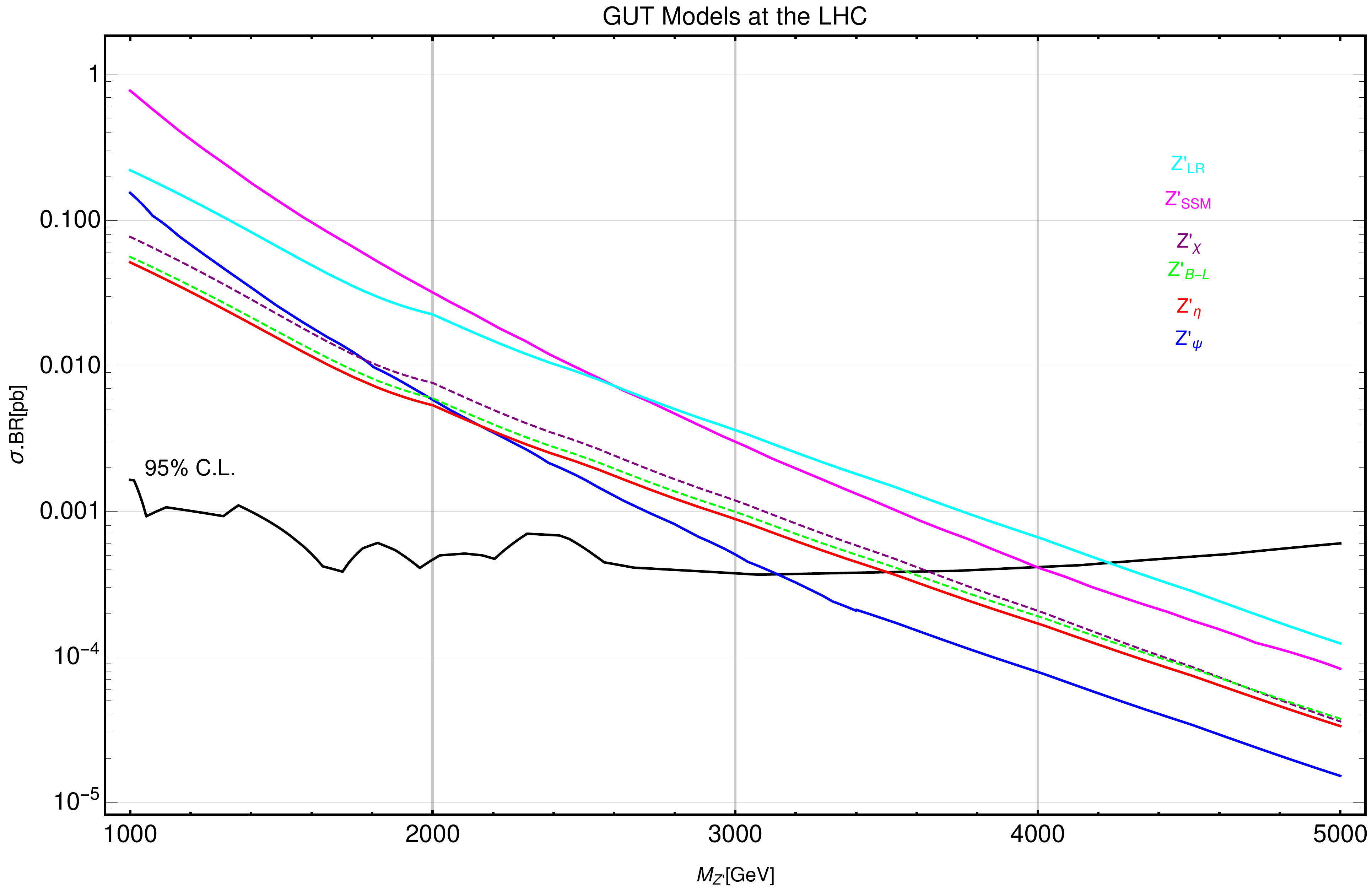}
\end{center}
\caption{Production cross section times branching ratio for the combined dilepton channel ($ee$+$\mu\mu$) at 13 TeV. The black curve is the 95\% C.L. limits from ATLAS using $13.3 ~\mathrm{fb^{-1}}$ of integrated luminosity. From {\it bottom to top} the curves delimit the results for $Z_\psi^\prime$, $Z_\eta^\prime$, $Z_{B-L}^\prime$, $Z_\chi^\prime$, $Z_{SSM}^\prime$ and $Z_{LR}^\prime$ models. A table with lower mass bounds for current and planned future hadron colliders can be found below .}
\label{fig:2}
\end{figure*}

Having described in detail all important ingredients for study, we can provide a more quantitative taste of dilepton searches. By looking back at Fig.\ref{fig:1} and taking the bin with invariant dielectron invariant mass of $3000-6000$~GeV, one can infer that with $13.3~ \mathrm{fb^{-1}}$ of integrated luminosity at 13 TeV around 100 events are expected from the B-L model. However, ATLAS collaboration has measured only $0.1\pm 0.026$ events \cite{ATLAS:2016cyf}. Since the new physics contribution far exceeds any statistical or systematic error, a reliable bound can be derived on the B-L model. A similar reasoning can be applied to all models under study. By repeatedly doing this comparison in a bin-by-bin model independent basis ATLAS collaboration placed 95\% C.L. limits on the underlying particle physics input quantity namely production cross section times branching ration into charged leptons (dielectron $+$ dimuon), represented by a black solid line in Fig.\ref{fig:2}. That said, we computed the production cross section for the six models under study following the receipt aforementioned and compared with the 95\% upper limit from ATLAS collaboration as shown in Fig.\ref{fig:2}. Our results for the SSM, $Z_\chi$ and $Z_\eta$ models agree well with the ones reported by ATLAS collaboration in \cite{ATLAS:2016cyf}, as well as with the analysis of \cite{Alves:2015mua,Klasen:2016qux} concerning the B-L model. See \cite{Accomando:2016sge,Cerrito:2016qig,Accomando:2016ouw} for other complementary studies of GUT models at the LHC.  

Moreover, we obtain projection sensitivities for $\sqrt{s}=33$ and $100$TeV energies having in mind the proposed proton-proton colliders, namely the high-energy LHC \cite{Avetisyan:2013onh,Cohen:2013xda,Apollinari:2015bam} and $100$~TeV collider \cite{Hinchliffe:2015qma,Arkani-Hamed:2015vfh,Contino:2016spe,Mangano:2016jyj,Goncalves:2017gzy}. The former is proposed to reach a center-of-energy of 33 TeV and up to $300 ~ \mathrm{fb^{-1}}$, whereas the latter is projected to reach from $\mathcal{L}\sim 1-30\, \mathrm{ab^{-1}}$ \cite{Hinchliffe:2015qma}. We conservatively adopt $\mathcal{L}=5\, \mathrm{ab^{-1}}$. In order to derive projected bounds we follow the recommendation of the CERN code that yields reasonable predictions \cite{Colliderreachcode}, and we solve for $M_\textrm{\small new}$ the equation, 

\begin{equation}
\frac{N_\textrm{\small signal events}(M^2_\textrm{\small new},E_\textrm{\small new},\mathcal{L}_\textrm{\small new})}{N_\textrm{\small signal events}(M^2,13~{\rm TeV},13.3 fb^{-1})}=1,
\end{equation}where M is the current bound on the $Z^{\prime}$ mass, and the number of events is estimated by computing the production cross section at a given center-of-energy with certain luminosity. This procedure has been validated in \cite{Rizzo:2015yha}, where the predictions for 13 TeV results agree well with experimental limits.

Our results are summarized in Table II. It is clear that a major sensitivity boost is expected when ramping up the center-of-energy from 14 TeV to 33 TeV, where $Z^{\prime}$ masses near 10 TeV become available. 
Furthermore, a 100 TeV collider with the modest luminosity of $5ab^{-1}$ is sensitivity to $Z^{\prime}$ masses between  $30-39$~TeV. Hopefully these discovery machines will be built and spot a signal at the multi-TeV scale \cite{Hinchliffe:2015qma,Arkani-Hamed:2015vfh,Baglio:2015wcg,Contino:2016spe,Mangano:2016jyj,Goncalves:2017gzy}. See \cite{Deppisch:2015qwa,Dev:2016dja,Chekanov:2016ppq,Kobakhidze:2016mdx,diCortona:2016fsn,Craig:2016ygr,Zhao:2016tai,Lillard:2016jxz,Antusch:2016nak,Englert:2017gdy} for other interesting sensitivity reach of a 100TeV collider.

In the next section, our results based on collider physics will be put into perspective with dark matter searches at direct detection experiments \footnote{We will ignore indirect dark matter detection limits \cite{Hooper:2012sr,Queiroz:2014zfa,Queiroz:2016gif}, as well as limits from flavor physics since for the models under study they are rather subdominant \cite{Queiroz:2014yna,Gonzalez-Morales:2014eaa,Allanach:2015gkd,Baring:2015sza,Queiroz:2015utg,Mambrini:2015sia,Alves:2015pea,Patra:2015bga,Queiroz:2016zwd,Profumo:2016idl,Campos:2017odj}.}.

\begin{table*}
\begin{tabular}{|c|c|c|c|c|c|c|c|c|}
\hline
{\color{blue} Model}  & {\rm $13$~TeV,$13.3~fb^{-1}$ } & {\rm $13$~TeV,$37~fb^{-1}$ } & {\rm $14$~TeV,$100~fb^{-1}$ } & {\rm $14$~TeV,$300~fb^{-1}$ } & {\rm $33$~TeV,$100~fb^{-1}$ } & {\rm $33$~TeV,$300~fb^{-1}$ }  &  {\rm $100$~TeV,$5~ab^{-1}$ }\\
\hline
{\color{blue} $Z^\prime_\psi$} & 3.13 TeV & 3.68 TeV & 4.46 TeV & 5.13 TeV & 7.98 TeV & 9.47 TeV & 30.54 TeV \\
\hline
{\color{blue} $Z^\prime_\eta$} & 3.47 TeV & 4.04 TeV & 4.85 TeV & 5.51 TeV & 8.85 TeV & 10.38 TeV & 33.25 TeV \\
\hline 
{\color{blue} $Z^\prime_{B-L}$ }& 3.55 TeV & 4.11 TeV & 5.55 TeV & 5.59 TeV & 9.03 TeV & 10.56 TeV & 33.8 TeV\\
\hline
{\color{blue} $Z^\prime_\chi$} & 3.63 TeV & 4.19 TeV & 5.55 TeV & 5.68 TeV & 9.23 TeV & 10.76 TeV & 34.41 TeV \\
\hline
{\color{blue} $Z^\prime_{SSM}$} & 4.02 TeV & 4.59 TeV & 6.05 TeV & 6.09 TeV & 10.21 TeV & 11.75 TeV & 37.36 TeV \\
\hline
{\color{blue} $Z^\prime_{LR}$} & 4.23 TeV & 4.8 TeV & 6.27 TeV & 6.31 TeV & 10.73 TeV & 12.28 TeV & 38.92 TeV \\
\hline
\end{tabular}
\caption{\footnotesize{Summary of current and projected bounds on the $Z^{\prime}$ mass in the SSM and various GUT models, in light of current and future proton-proton colliders.}}
\label{tab:Zpcouplings}
\end{table*}

\section{Connection to Dark Matter} 

The nature of dark matter is one of the most fascinating puzzle in science \cite{Queiroz:2016sxf,Bertone:2016nfn}. In order to unveil its nature it is desirable to collect data across different but complementary search strategies, such as collider and direct detection. Vector mediators are a special example in this direction since both collider and direct detection observables are strongly dictated by the $Z^{\prime}$ properties. In particular, a $Z'$ boson represents an attractive portal for interaction within the WIMP paradigm \footnote{Viable DM can be accommodated in GUT frameworks also without relying on the WIMP paradigm~\cite{Mambrini:2013iaa,Mambrini:2015vna}}. In a GUT inspired framework the DM would be represented by a new particle state belonging to a suitable representation such that it has not trivial quantum numbers with respect to the symmetry group associated to the $Z'$ while being singlet with respect to the SM group\footnote{Notice that, in general, in this kind of construction the DM is actually part of a multiples so other states might be relevant for its phenomenology.}. Interestingly the stability of the DM does not require the imposition of ad hoc discrete (or global) symmetries for its stability, as customary done in simplified realizations~\cite{Abdallah:2015ter,Frandsen:2012rk,Alves:2013tqa,Alves:2015mua,Alves:2016cqf}, since they might naturally arise as remnants in the different breaking steps of the GUT group~\cite{Mambrini:2015vna}~\footnote{See e.g. here~\cite{Mambrini:2015sia,Gross:2015cwa,Arcadi:2016kmk,Alves:2016fqe} for alternative examples of natural emergence of DM stability.}. The eventual presence of a DM candidate might have a sizable impact on collider phenomenology. Indeed, in case the $Z' \rightarrow \textrm{DM DM}$ decay process is kinematically allowed, a sizable invisible branching ratio would weaken the limit from searches of dilepton resonances~\cite{Arcadi:2014lta} since the corresponding cross section is reduced by a factor $1-\textrm{BR}( Z' \rightarrow \textrm{DM DM})$. This creates an interesting complementarity with Dark Matter searches as well as the DM relic density, since they constraint the possible values of $\textrm{BR}(Z' \rightarrow \textrm{DM DM})$. LHC limits have, in turn, impact on the DM phenomenology. For example, too strong limits on the mass of the $Z'$ would correspond in general to a suppressed pair annihilation cross section, hence implying an overabundant DM~\footnote{This issue could be overcome in the case that additional particle states influence DM relic density, e.g. through coannihilations, or by invoking non-thermal DM production~\cite{Arcadi:2011ev,Chu:2013jja} or more generally, modified cosmological histories~\cite{DEramo:2017gpl}.} (see next section for more details).

\begin{figure}[htb]
\begin{center}
\includegraphics[width=8 cm]{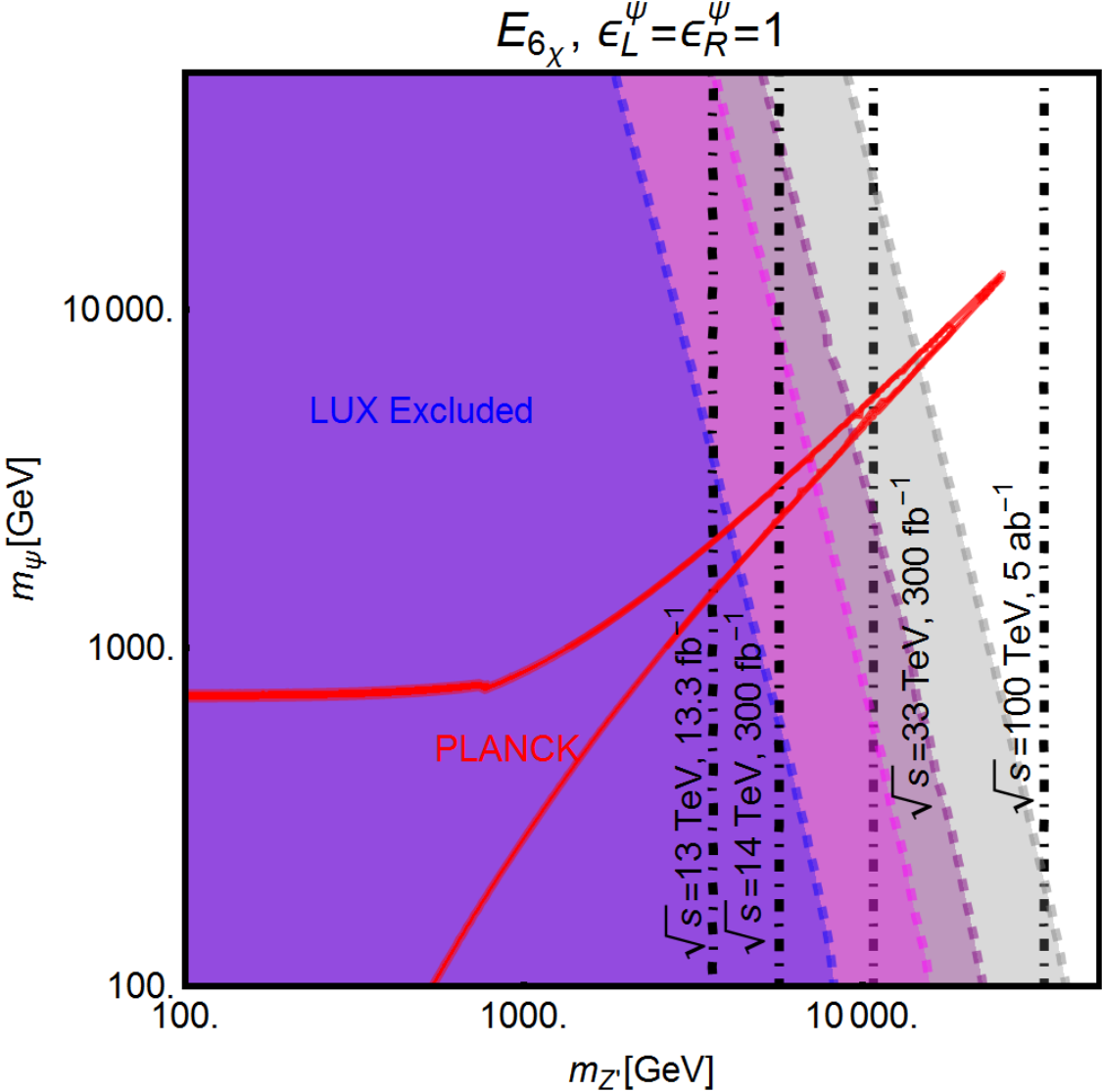}
\end{center}
\caption{\footnotesize{Comparison between DM current/projected constraints and current/projected constraints on the mass of the $Z'$ from collider searches. The DM has been chosen to be a dirac fermion and the couplings of the $Z'$ with the SM fermions are dictated by the $E_{6,\chi}$ model. In the plot the red line represents the isocontour of the correct DM relic density. The region at the left of the blue dashed line is ruled out by DD constraints by LUX~\cite{Akerib:2016vxi}. Regions at the left of the magenta, purple and gray dashed lines correspond, respectively, to the projected sensitivities of XENON1T~\cite{Aprile:2015uzo}, LZ~\cite{Szydagis:2016few} and Darwin~\cite{Aalbers:2016jon}. The black lines represent current (first line on the left) and projected exclusions by LHC of dilepton resonances (the corresponding values of center of mass energy and luminosity are reported in vicinity of the lines). The region at the left of each line should be regarded as experimentally ruled out in case no signal is detected at the values of center of mass energy and luminosity reported in proximity of the line itself.}}
\label{fig:DM}
\end{figure}

We have shown, in fig.~\ref{fig:DM}, an example of this kind of complementarity (this topic has been more extensively reviewed e.g. in~\cite{Arcadi:2017kky}). 

We have focused here on the case of a Dirac DM candidate $\psi$ coupled to the $Z'_\chi$. The relevant Lagrangian for DM interactions can be written in analogous way as the one of the SM fermions:  

\begin{equation}
\mathcal{L}= g_{f} \bar \psi \gamma^\mu \left(\epsilon^{\psi}_L P_L + \epsilon^{\psi}_R P_R\right) \psi Z^{'}_\mu
\end{equation} 

We have set for simplicity $\epsilon^{\psi}_L=\epsilon^{\psi}_R=1$ (this choice is actually rather special since would imply only vectorial coupling of the DM with the $Z'$ but it is nevertheless not problematic given the illustrative purposes of the fig.~\ref{fig:DM}~\footnote{Notice also that an axial coupling of the DM with the $Z'$ \cite{,Lebedev:2014bba} could potentially lead to violation of unitarity from the annihilation process $\psi \psi \rightarrow Z'Z'$.~\cite{Kahlhoefer:2015bea}. This problem is automatically cured in UV complete frameworks, like a GUT theory, by the presence of the Higgs fields responsible of the breaking of the $U(1)$ gauge symmetry associated to the $Z'$. A proper treatment would then require an explicit construction of the GUT model, which is not in the purpose of this letter.})

For what regards the DM phenomenology we have required the correct DM relic density according the WIMP paradigm, i.e. $\Omega h^2 \propto 1/\langle \sigma v \rangle$ where $\langle \sigma v \rangle$ is the thermally averaged DM pair annihilation cross-section. The experimentally favored value $\Omega h^2 \approx 0.12$~\cite{Ade:2015xua} corresponds to an annihilation cross-section of the order of $10^{-26}{\mbox{cm}}^3 {\mbox{s}}^{-1}$ (for details on the quantitative determination of the relic density we refer to~\cite{Arcadi:2017kky}). Regarding DM searches we have focused on Direct Detection, which provides the strongest constraints in the considered scenarios, which relies on Spin Independent (SI) interactions of the DM with nucleons which are described by a cross-section of the form (for definiteness we consider the case of scattering on protons):

\begin{equation}
\sigma_{\psi,p}^{\rm SI}=\frac{g_f^4 \mu_{\psi p}^2}{\pi m_{Z'}^4}V^2_\psi{\left[f_p \frac{Z}{A}+f_n \left(1-\frac{Z}{A}\right)\right]}^2 \nonumber\\
\end{equation}

\noindent
where $f_p=2 V_u+V_d,\,\,\,\,f_n=V_u+2 V_d$, $\mu_{\psi p}$ is the DM-proton reduced mass while $Z$ and $A$ represent the number of protons and the total number of nucleons of the detector material (we will consider Xenon--type detectors). The parameters $V_{\psi,u,d}$ in the equation above represent the vectorial coupling of the DM, up and down quarks to the $Z'$, i.e. $V_{f=u,d,\psi}=\frac{1}{2}(\epsilon_L^f+\epsilon_R^f)$.

Fig.~\ref{fig:DM} reports in a bidimensional plane of the DM and $Z'$ masses the curve of the correct DM relic density, and the current limits, provided at present times by the LUX experiment~\cite{Akerib:2016vxi}, and projected limits next future experiments. For these we have considered the Xenon1T~\cite{Aprile:2015uzo}, LZ~\cite{Szydagis:2016few} and Darwin~\cite{Aalbers:2016jon} experiments. Notice that the sensitivity of the Darwin experiment is comparable to the expected value of the cross-section associated to coherent neutrino scattering on nuclei, which represents somehow the ultimate reach of experiment probing elastic scattering of WIMPs on nuclei. The DM constraints have been compared with the current LHC exclusion limit~\footnote{For parameters adopted in the analysis the invisible branching fraction of the $Z'$ is typically rather small and the impacts in a negligible way limits from resonance searches.} from dilepton resonance searches as well as the maximal reaches for the three values of the center of mass energy considered in this work (see tab.~\ref{tab:Zpcouplings}). As evident future collider limits can overcome current and future limits by Direct Detection. For the chosen assignation of the parameters, an absence of signals at a 100 TeV collider would completely rule out the WIMP hypothesis in this framework.

\section{Conclusions} 

Grand Unified Theories (GUT) provide an unified description of electromagnetic, weak and strong interactions at high energy scales around $10^{16}$ GeV. These kind of theories can be nevertheless probed through collider studies in the case the GUT gauge symmetry group is broken, before EW symmetry breaking, in some subgroup larger than the SM gauge group. A phenomenological and gripping method to probe this kind of scenario consist in to study Abelian groups, which can be embedded in GUT frameworks, which thus predict the existence of a new neutral gauge boson, a $Z^{\prime}$, whose couplings with the SM fermions are dictated by the breaking patter of the GUT group itself. The observation of a signal of a $Z^{\prime}$ at TeV scale, with interactions strength as predicted by GUT-inspired models, would then constitute a hint of GUT at high energy scales. 
In light of the current null results, we used up-to-date dilepton data from LHC to derive  the lower mass bounds for several GUT models.  Moreover, we casted projected limits having in mind possible future colliders namely, the high-energy LHC and $100$~TeV collider, with the latter being able to probe $Z^{\prime}$ masses around $38$~TeV. 
Lastly, we put our findings into the perspective of a connection with the DM problem. Interpreting the $Z^{\prime}$ at the mediator (portal) of the interactions between the DM and the SM fermions, we have exploited, in a simple example with dirac fermion DM, the complementarity between collider searches and DM direct detection experiments. 

\section*{Acknowledgements}
The authors warmly thank Alexandre Alves, Carlos Yaguna, Werner Rodejohann for fruitful discussions. This work is also supported by the Spanish MICINN's Consolider-Ingenio 2010 Programme under grant Multi-Dark 
{\bf CSD2009-00064}, the contract {\bf FPA2010-17747}, the France-US PICS no. 06482 and the LIA-TCAP of CNRS. 
Y.~M. acknowledges partial support the ERC advanced grants Higgs@LHC and MassTeV. 
This research was also supported in part by the Research Executive Agency (REA) of the European Union under
the Grant Agreement {\bf PITN-GA2012-316704} (``HiggsTools'').



\bibliography{darkmatter}

\end{document}